\documentclass[a4paper, aps, prd, longbibliography, preprint, superscriptaddress,nofootinbib]{revtex4-1}

\usepackage{asymptote} 
\usepackage{array}
\usepackage{amsmath,amssymb,amsfonts}
\usepackage[colorlinks]{hyperref}
\usepackage{multirow}
\usepackage{float} 
\usepackage[utf8]{inputenc}
\usepackage{graphicx}
\usepackage{subfigure}
\graphicspath{{./pic/}}

\usepackage[utf8]{inputenc}
\newcommand{\bw}{\begin{widetext}}
\newcommand{\ew}{\end{widetext}}
\newcommand{\be}{\begin{equation}}
\newcommand{\en}{\end{equation}}
\newcommand{\bee}{\begin{equation}}
\newcommand{\ene}{\end{equation}}
\newcommand{\bea}{\begin{eqnarray}}
\newcommand{\ena}{\end{eqnarray}}
\newcommand{\bes}{\begin{subequations}}
\newcommand{\ens}{\end{subequations}}
\newcommand{\bef}{\begin{figure}}
\newcommand{\enf}{\end{figure}}

\def\ie{{\it i.e.}}
\def\etc{{\it etc.~}}

\def\cald{\mathcal{D}}

\def\cali{\mathcal{I}}

\def\calk{\mathcal{K}}
\def\call{\mathcal{L}}
\def\calm{\mathcal{M}}
\def\caln{\mathcal{N}}

\def\calp{\mathcal{P}}

\def\to{\rightarrow}

\def\tev{{\rm TeV}}
\def\gev{{\rm GeV}}
\def\mev{{\rm MeV}}

\begin{document}


\title{Polarization and Correlation Effects in Lepton Flavor Violated Decays Induced by Axion-Like Particle}

\author{Kai Ma}
\email[Electronic address: ]{makai@ucas.ac.cn}
\affiliation{School of Fundamental Physics and Mathematical Science, Hangzhou Institute for Advanced Study, UCAS, Hangzhou 310024, Zhejiang, China}
\affiliation{International Centre for Theoretical Physics Asia-Pacific, Beijing/Hangzhou, China}
\affiliation{Department of Physics, Shaanxi University of Technology, Hanzhong 723000, Shaanxi, China}

\date{\today}

\begin{abstract}
Lepton flavor violated processes are strongly suppressed in the Standard Model, 
but can be sizable in some extended models. 
The current experimental bounds on decay modes $\ell'^{\pm} \to a \ell^{\pm}$ 
are much weaker than the other processes because of 
the huge irreducible backgrounds $\ell'^{\pm} \to \ell^{\pm} \bar{\nu}_{\ell} \nu_{\ell'}$.
In this paper, we study polarization effects of both the signals and backgrounds.
We find that signals and backgrounds have distinctive polarization effects, 
and for the irreducible backgrounds both longitudinal and transverse 
polarization effects survive when the relative momentum of the two neutrinos 
are integrated out. At low energy $e^+e^-$ collider, for instance 
the Belle II experiment, leptons are generated in pair 
but with no net polarization. However, we show that polarization correlation of 
the lepton pair is a useful observable for probing the signal.
More interestingly, the polarization correlation depends on product of 
scalar and pseudo-scalar couplings, and hence are sensitive to their relative sign.
Because kinematical reconstruction or tagging is not necessary,
number of the available signal events increases significantly.
\end{abstract}

\maketitle

\tableofcontents
\newpage

\section{Introduction}
It is generally accepted that the Standard Model (SM) needs to be improved 
to account for new physics (NP) phenomenologies, 
for examples neutrino oscillations and Dark Matter \etc. 
Usually, new heavy particles are introduced in beyond SMs (BSMs), 
and high energy colliders are expected to be efficient for detecting those heavy particles.
However, so far no convincing evidence of new particle has been observed,
for instance, at the Large Hadron Collider (LHC). On the other hand, 
light or ultra light particles can also naturally appear as pseudo-Nambu-Goldstone 
bosons of an underlying broken symmetry. A promising instance is the 
Axion model~\cite{Peccei:1977hh,Weinberg:1977ma,Wilczek:1977pj}, which was originally 
introduced to solve the strong CP problem, but was extended effectively to be correlated 
with other fundamental open questions in particle physics, such as origin of the
dark matter (DM)~\cite{Preskill:1982cy,Abbott:1982af,Dine:1982ah,Grange:2015fou}, 
as well as the hierarchy~\cite{Dvali:2003br,Graham:2015cka} and flavor 
problems~\cite{Wilczek:1982rv,Davidson:1981zd,Roy:1981me,Ema:2016ops,Calibbi:2016hwq}.
Such kind of new states are generally referred to as 
Axion-Like-Particles (ALPs)~\cite{Rubtsov:2014uga,Arvanitaki:2009fg,Irastorza:2018dyq}.
In view of effective field theory (EFT), mass of the ALP and couplings to the SM particles 
are free parameters to be determined experimentally~\cite{Georgi:1986df}.
Stringent constrains on ALP couplings arise from cosmological and astrophysical 
observations~\cite{Rubtsov:2014uga,Arvanitaki:2009fg,Irastorza:2018dyq,DiLuzio:2016sbl,Georgi:1986df,Berezhiani:1990wn,Berezhiani:1990jj,Berezhiani:1989fp,Sakharov:1994pr}.
The LEP collaboration studied ALP, and obtained significant bounds on its masse 
ranging from the $\mev$ scale up to $90\gev$~\cite{Jaeckel:2015jla}.
Searches for ALP at the LHC and future colliders have also been extensively explored
~\cite{Jaeckel:2015jla,Knapen:2016moh,Flacke:2016szy,Brivio:2017ije,Feng:2018pew,Bauer:2018uxu,CidVidal:2018blh,Harland-Lang:2019zur,Aloni:2019ruo,Gavela:2019cmq,Bauer:2017ris}. 
Furthermore, decays of mesons at flavor factories can also provide rich possibilities 
to probe ALP couplings to both leptons and quarks~\cite{Dolan:2014ska,Dolan:2017osp,Merlo:2019anv}.

More interestingly, while lepton flavor changing neutral currents in the SM
are strongly suppressed by the neutrino mass-squared 
differences~\cite{Petcov:1976ff,Hernandez-Tome:2018fbq}, 
but can be effectively violated at tree-level by the 
ALPs~\cite{Batell:2009jf,Freytsis:2009ct,Izaguirre:2016dfi,Altmannshofer:2019yji,
Albrecht:2019zul,Gavela:2019wzg,Dobrich:2018jyi}.
Recent studies showed that large parameter space of flavor violated ALP 
can offer qualitatively new ways to explain anomalies related to the 
magnetic and electric dipole moments of the muon and 
electron~\cite{Bauer:2019gfk,Escribano:2020wua,Endo:2020mev,Iguro:2020rby}.
However, it was shown that in case of $m_{a} > m_{\tau}$, 
a simultaneous explanation of both anomalies in terms of flavor-violating ALP couplings to 
$\tau$-lepton is not possible due to relatively large contribution to $\mu \to e \gamma $, 
unless there is a large hierarchy between couplings of electron and muon to the 
$\tau$-lepton~\cite{Bauer:2019gfk}. Similar situation happens in the case of
$m_{\mu} < m_{a} < m_{\tau}$. In case of that $2m_{e} < m_{a} < m_{\mu}$, 
flavor diagonal contributions have to be dominant in order to account for 
these two anomalies simultaneously, and the ALP couplings to electron
should be larger than the ALP couplings to muon with opposite sign~\cite{Bauer:2019gfk}.
If the ALP is ultra light, flavor-violating ALP couplings can never give correct contributions 
to the anomalies, and hence can put strong limits on the corresponding 
couplings~\cite{Coriano:2021pjr,Hsu:2020ikn,Brzeminski:2020uhm,Ng:2020ruv,
Norton:2020ert,Rogers:2020ltq,Dev:2020kgz,Gonzalez:2020fdy}.
In this case, flavor violated decay modes $\tau \to \ell a$ can provide direct 
measurement on the relevant parameters. 
However, It is challenging to distinguish signals from the irreducible backgrounds,
$ \tau^{\pm} \to \ell^{\pm} \bar{\nu}_{\ell}\nu_{\tau}$, 
where more than one neutrino appear in the final state.

The ARGUS collaboration studied these channels,
and an upper limit $\sim 2\%$ was obtained by tagging 
one side of the $\tau$-lepton pair decaying in 3-prong mode~\cite{Albrecht:1995ht}. 
The $3-1$ prong searching method was also used by the Belle II 
experiment~\cite{Abe:2010gxa,Villanueva:2018pbk,Konno:2020+A,Kou:2018nap}.
However, number of events of the signal was significantly reduced 
because of small branching fraction of the 3-prong decay channel. 
Some new methods were
proposed~\cite{DeLaCruz-Burelo:2020ozf,Xiang:2016jni,Christensen:2014yya}, 
but still rely on double-side tagging of the $\tau$-lepton.
In this sense, measuring total cross section and/or decay rates are 
ineffective due to one-side tagging as well as the huge irreducible background.
For decay of the $\mu$-lepton, even through mass spectrum of the 
invisible particle(s) (two neutrinos or the ALP) can be measured, 
but the signal process  $\mu^{\pm} \to a e^{\pm}$ can be strongly suppressed 
because the muon beam is highly polarized in the direction opposite to the muon 
momentum~\cite{Yamanaka:1986ze,Baldini:2020okg,Hirsch:2009ee,Anselm:1985bp,
Sanchez-Glez:2020sgh,Aguilar-Arevalo:2020ljq,Andreev:2006wh,Gordeev:2002gi}.
This is also the reason of that experimental bounds on the decay modes 
$\ell^{'\pm} \to a \ell^{\pm} $ are much weaker than the other flavor violated 
processes~\cite{Cornella:2019uxs}.

Alternatively, polarization effects can provide more efficient observables 
for measuring the flavor violation parameters. It is well-known that, 
longitudinal polarization of a $\tau$-lepton can be efficiently measured 
by using its decay products~\cite{Tsai:1971vv,Kuhn:1982di,Nelson:1988bz,Nelson:1989cy,
Bernreuther:1989kc,Bernabeu:1989ct,Bernabeu:1990na,Bullock:1991my,Bullock:1992yt}.
The hadronic decay mode 
$\tau \to \pi \nu_{\tau}$ has maximum sensitivity~\cite{Tsai:1971vv,Kawasaki:1973hf}.
Sensitivities of the other hadronic decay modes, $\tau \to \rho \nu_{\tau}$ and 
$\tau \to a_{1} \nu_{\tau}$, are relatively lower due to deconstructive contributions from 
different helicity amplitudes, but can be significantly improved by further kinematical
selections of the helicity components via subsequent decays of the
$\rho$-meson~\cite{Rouge:804793,Rouge:1990kv,Hagiwara:1989fn} 
and $a_{1}$-meson~\cite{Hagiwara:1989fn}, respectively.
However, for the leptonic decay modes,
$ \tau^{\pm} \to \ell^{\pm} \bar{\nu}_{\ell}\nu_{\tau}$,
since two neutrinos appear in the final state,
polarization effects are weaker than the other modes.
Because of this, searching for new physics by employing polarization effects 
of the leptonic decay modes was usually not considered as useful channels.

In this paper we study polarization effects of lepton flavor violated decays 
$\ell'^{\pm} \to a \ell^{\pm} $ and the corresponding irreducible 
backgrounds $ \ell^{\pm} \to \ell^{\pm} \bar{\nu}_{\ell}\nu_{\tau}$.
In Sec.~\ref{sec:lfv:alp}, we show our parameterizations of the ALP model.
In Sec.~\ref{sec:LepDecay:Axion} and Sec.~\ref{sec:AxionDecay},
we study polarization effects of the decay processes $\ell'^{\pm} \to a \ell^{\pm} $ 
and $\ell'^{\mp} \to \ell^{\mp} \imath^{\pm} \jmath^{\mp}$, respectively.
In Sec.~\ref{sec:TauLepDecay}, we will show that polarization effects 
of the irreducible backgrounds survives when the invariant mass of the 
two neutrinos is inclusively measured. 
In case of that the $\ell'$-leptons are unpolarized,
for instance the $\tau$-lepton pair produced at the BelleII experiment,
we will show in Sec.~\ref{sec:correlations} that
polarization correlations of a $\ell'$-lepton pair are powerful for probing the signals. 
Summary of our studies are given in Sec.~\ref{sec:conclusion}.

\section{Lepton Flavor Violated ALP}\label{sec:lfv:alp}
In this section we give out our parameterizations of a lepton flavor violated model 
with a new scalar or pseudo scalar mediator, denoted by $a$. Up to dimension five, 
the relevant low-energy effective interaction Lagrangian of an ALP 
coupling to charged leptons and photons can be described 
by~\cite{Bauer:2019gfk,Cornella:2019uxs,Georgi:1986df},
\bea
\call^{\rm Eff}
&=&
\frac{1}{2} \big[ (\partial_{\mu} a)^2 - m_{a}^{2} a^{2} \big]
+
\call^{\rm Eff}_{\rm a\ell\ell'} + \call^{\rm Eff}_{\rm a\gamma\gamma'} \,,
\\[3mm]
\call^{\rm Eff}_{\rm a\ell\ell'} 
&=&
- \frac{1}{f_{a}} (\partial_{\mu} a) \left(\,
\overline{\ell_{L}} \, U_{L} \gamma^{\mu} \ell_{L} + 
 \, \overline{\ell_{R}} \, U_{R} \gamma^{\mu}\ell_{R} \right) \,,
\\[3mm]
\call^{\rm Eff}_{\rm a\,\gamma\gamma} 
&=&
\frac{\alpha_{E}}{4\pi} \frac{1}{f_{a}}  c_{\gamma\gamma}\, a F^{\mu\nu} \widetilde{F}_{\mu\nu}\,,
\ena
where $\ell = (e, \mu, \tau)^{T}$, the hermitian matrices $U_{L} $ and $U_{R}$ 
are defined in mass eigenstates of the leptons, 
and $f_{a}$ is the decay constant of the ALP. 
Since we are considering an ALP as a pseudo-Nambu-Goldstone boson of a 
spontaneously broken global symmetry, the decay constant $f_{a}$ characterizing 
the broken scale is naturally much larger than the low energy scale parameter, 
$m_{a}$, mass of the ALP. In case of that the ALP couples to on-shell leptons, 
by using equations of motion of the leptons, 
the above Lagrangians can be rewritten as,
\bee\label{eq:Larg:Aff}
\call^{\rm Eff}_{\rm a\ell\ell'} 
=
\frac{i}{f_{a}} a \left[\, U_{V}^{ij} ( m_{i} - m_{j} )
\overline{\ell^{i}} \, \ell^{j} + 
U_{A}^{ij} ( m_{i} + m_{j} ) \overline{\ell^{i}} \gamma^{5} \ell^{j} \right] \,,
\ene
where $U_{V}=(U_{R} + U_{L})/2$ and $U_{A}=(U_{R} - U_{L})/2$, and $m_{i}$ are masses 
of the leptons $\ell_{i}$. For flavor-diagonal ALP couplings, 
one can easily find that only the pseudo-scalar Lorentz structure survives. 
Since we will mainly consider on-shell leptons, 
the Lagrangian \eqref{eq:Larg:Aff} is hence the relevant part studied in this paper.
Furthermore, we introduce dimensionless scalar couplings, 
and the Lagrangian \eqref{eq:Larg:Aff} is rewritten as,
\bee\label{eq:Model:Lep:Sim}
\call^{\rm Eff}_{\rm a\ell\ell'} 
=
- \, a \,\overline{\ell^{i}} \left( S_{S}^{ij} +  S_{P}^{ij} \gamma^{5}  \right)  \ell^{j} \,,
\ene
where the scalar and pseudo-scalar coupling matrices are given as,
\bee\label{eq:mCouplings}
S_{S}^{ij} = \frac{i}{f_{a}}  U_{V}^{ij} ( m_{j} - m_{i}  ) \,,
\;\;\;
S_{P}^{ij}  = - \frac{i}{f_{a}}  U_{A}^{ij} ( m_{i} + m_{j} ) \,.
\ene
Here we have used different parameterizations such that the diagonal elements 
have the same sign conventions with the standard Yukawa couplings.
These two kinds of conventions can have (sign) differences in some observables, 
particularly when we consider interference effects between SM and Axion induced channels.
Sometimes we also use following notations with explicit chirality,
\bee
S_{L}^{ij} = S_{S}^{ij} -  S_{P}^{ij} \,,
\;\;\;
S_{R}^{ij} = S_{S}^{ij} +  S_{P}^{ij} \,.
\ene

Phenomenologies details of the above ALP model strongly depends on its mass.
In case of that the ALP is ultralight, star evolution can be significantly affected 
as a result of stellar cooling mechanism~\cite{Raffelt:1994ry}, 
and hence can place strong constraints on the ALP couplings by 
astrophysical observations~\cite{DiLuzio:2020wdo,Calibbi:2020jvd,Bollig:2020xdr,
Croon:2020lrf,DiLuzio:2020jjp}.
A very strong bound on the pseudo-scalar coupling of electron, 
$S_{P}^{ee}<2.1\times10^{-13}$ at 90\% C.L., was obtained in Ref.~\cite{Calibbi:2020jvd}.
Similar constraint on theu $\mu$-lepton, $S_{P}^{\mu\nu}<2.1\times10^{-10}$, 
was reported in Ref.~\cite{Croon:2020lrf}. 
For the corresponding scalar couplings, $S_{S}^{\ell\ell}$,
similar upper bounds applies as argued in Ref.~\cite{Escribano:2020wua}. 
Here we further assume that those constrains are also valid as long as $m_{a} < 2 m_{e}$.
It turns out that a light ALP can couple to leptons only through flavor violated couplings. 
For $\mu-e$ flavor violated couplings, 
a conservative bound $S_{S/P}^{e\mu}<5.3\times10^{-11}$
was obtained~\cite{Hirsch:2009ee,Bayes:2014lxz,Calibbi:2020jvd}.
Substantially improvement on this coupling will be reached by the Mu3e experiment 
in near future~\cite{Calibbi:2020jvd,Perrevoort:2018ttp}. For the $\tau$-lepton, 
the currently best experimental limits $S_{S/P}^{\tau\ell} \lesssim 10^{-7} $, 
which are relatively weak compared to the couplings $S_{S/P}^{e\mu}$,
were obtained by the ARGUS collaboration~\cite{Albrecht:1995ht}.

\section{Polarized Effects and Angular Distributions}
\label{sec:pol}

\subsection{Polarization Effects in Decays $\ell' \to \ell a$}
\label{sec:LepDecay:Axion}
In this subsection we study polarization effects and the corresponding 
angular distributions of following decay process,
\bee
\ell'^{-}( \vec{p}_{\ell'}, \lambda_{\ell'}  ) 
\longrightarrow 
a(\vec{p}_{a} )
+
\ell^{-}(\vec{p}_{\ell}, \lambda_{\ell} ) \,.
\ene
Here $\lambda_{\ell^{\prime}}$ and $\lambda_{\ell}$ are helicities of 
the decaying and outgoing leptons, $\ell'$ and $\ell$, respectively. 
In principal, the ALP can undergo further transitions.
However, if decay width of the ALP is very small, 
it may not decay inside of the detector, and hence is effectively invisible. 
We will study polarization effects of this process in Sec.~\ref{sec:AxionDecay}. 
For this moment let us focus on the (effectively) invisible decay $\ell' \to \ell a$.

The helicity amplitudes as well as the corresponding helicity density matrix elements 
are calculated in the rest frame of the lepton $\ell'$. And the $z$-axis is defined as the
flying direction of the lepton $\ell'$. For this moment we assume that the $x$-axis 
can be defined unambiguously such that the helicity density matrix elements are also
non-trivial functions of the azimuthal angle 
(once production dynamics of the $\ell'$-lepton is given, 
the $x$-axis can be defined as a direction lying the scattering plane). 
Parameterizations of the related momentum and helicity amplitudes given in App.~\ref{app:kin:LTA}.
The diagonal and off-diagonal elements of the decay density matrix elements are given as,
\bea
\cald_{\lambda_{\ell'}}^{ \lambda_{\ell'} }  
&=&
\frac{1}{2}m_{\ell'}^{2} \alpha_{a\ell}^{+} \alpha_{a\ell}^{-} \, S_{\ell \ell'}^{+}
\left[
1 + \gamma_{\ell}^{-1} \kappa_{\ell \ell'}^{-}  
+
\lambda_{\ell' } \xi_{\ell\ell'}^{+} \beta_{\ell} \cos\theta_{a}  \right]\,,
\\[3mm]
\cald_{\lambda_{\ell'}}^{-\lambda_{\ell'}}  
&=&
\frac{1}{2} m_{\ell'}^{2}\alpha_{a\ell}^{+} \alpha_{a\ell}^{-}  \,S_{\ell\ell'}^{+} \, 
\xi_{\ell\ell'}^{+} \, e^{i\lambda_{\ell'}\phi_{a}} \sin\theta_{a}  \,,
\ena
where the overall normalization constant $S_{\ell \ell'}^{+}$ is given as 
$S_{\ell \ell'}^{\pm} = \big|S_{S}^{\ell \ell'}\big|^2 \pm \big|S_{P}^{\ell \ell'}\big|^2$, and
the effective coupling constants $\kappa_{\ell\ell'}^{\pm}$ and $\xi_{\ell\ell'}^{+}$ are defined as follows,
\bea
\kappa_{\ell\ell'}^{\pm} 
&=&  
\left( \big|S_{S}^{\ell \ell'}\big|^2 \pm \big|S_{P}^{\ell \ell'}\big|^2  \right) /S_{\ell \ell'}^{+} \,,
\\[3mm]
\xi_{\ell\ell'}^{+} &=&  2\Re\left\{ S_{S}^{\ell\ell'} S_{P}^{\ell \ell'\ast}\right\} / S_{\ell \ell'}^{+} \,.
\ena

Since the transverse polarization, which is represented by the azimuthal angle
dependences, can be useful if the rest frame of the $\ell'$-lepton can be 
reconstructed (as we assumed in the above equations). However, since
$a$ is invisible it is usually challenging. 
So here we discuss only the longitudinal polarization effects.
In term of polar angle of the charged lepton $\ell$, $\theta_{\ell} = \pi -\theta_{a}$,
differential decay width of a polarized $\ell'$-lepton is given as,
\bee\label{eq:Axion:invDW}
\frac{1}{ m_{\ell'} }\frac{ d \Gamma_{ \lambda_{\ell'}  } }{ d \cos\theta_{\ell}  }
=
\frac{ \overline{\beta}_{\ell'\ell a}  }{ 64 \pi }  \alpha_{a\ell}^{+} \alpha_{a\ell}^{-}  S_{\ell \ell'}
\left[
1 + \gamma_{\ell}^{-1} \kappa_{\ell \ell'}^{-}  
-
\lambda_{\ell' } \xi_{\ell\ell'}^{+} \beta_{\ell} \cos\theta_{\ell}  \right] \,.
\ene
We can clearly see that as long as $\xi_{\ell\ell'}^{+} \neq 0 $, \ie, both $S^{\ell\ell'}_{S}$
and $S^{\ell\ell'}_{P}$ have non-zero real components, 
then there is a nontrivial polarization effect. 
Most importantly, the polarizer $\xi_{\ell\ell'}^{+}$ is sensitive to 
the relative sign between $S^{\ell\ell'}_{S}$ and $S^{\ell\ell'}_{P}$. 
However as we have mentioned, because $a$ is invisible, 
kinematics of the final state usually can not be known precisely.  
Hence in practice it is challenging to directly employ the above polarization effect. 
However, energy fraction of the $\ell$-lepton,  $z_{\ell} = E_{\ell}/E_{\ell'}$, 
is correlated to the polar angle of the $\ell$-lepton.
At $e^{+}e^{-}$ collider it can be measured precisely, 
and hence is an excellent observable.
For a mother lepton $\ell'$ with a boost factor $\beta_{\ell'}$ in the Lab. frame, 
the energy fraction $z_{\ell}$ is given as,
\bee
z_{\ell}  = \frac{1}{2} \alpha_{a\ell}^{-} \left(  1 + \beta_{\ell'}\beta_{\ell}\cos\theta_{\ell} \right)\,.
\ene
In term of the energy fraction $z_{\ell}$, the differential decay width is given as,
\bee\label{eq:Axion:invDW:eFrac}
\frac{1}{ m_{\ell'} }\frac{ d \Gamma_{ \lambda_{\ell'}  } }{ d z_{\ell}  }
=
\frac{ \overline{\beta}_{\ell'\ell a}  \alpha_{a\ell}^{+}   }{ 16 \pi \beta_{\ell'}\beta_{\ell} }    S_{\ell \ell'}
\left[
\frac{1}{2} 
\left( 1 + \gamma_{\ell}^{-1} \kappa_{\ell \ell'}^{-} + 
\lambda_{\ell' } \xi_{\ell\ell'}^{+} \beta_{\ell'}^{-1} \alpha_{a\ell}^{-} \right)
-
\lambda_{\ell' } \beta_{\ell'}^{-1} \xi_{\ell\ell'}^{+} z_{\ell}  \right] \,,
\ene
\begin{figure}[ht]
\begin{center}
\subfigure[]{
\label{fig:Axion:eFrac:pXi}
\includegraphics[width=0.312\linewidth]{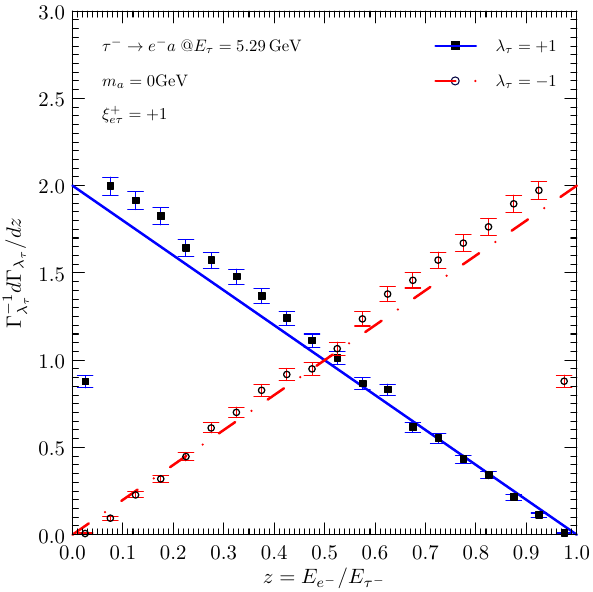}
}
\subfigure[]{
\label{fig:Axion:eFrac:nXi}
\includegraphics[width=0.312\linewidth]{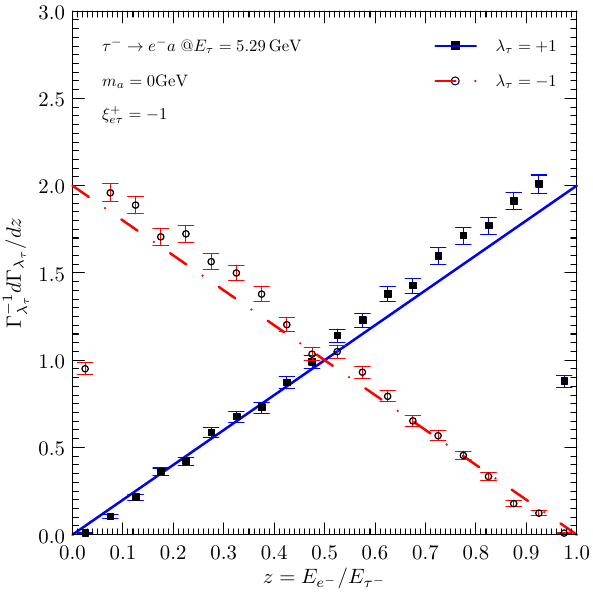}
}
\subfigure[]{
\label{fig:Axion:Gamma}
\includegraphics[width=0.312\linewidth]{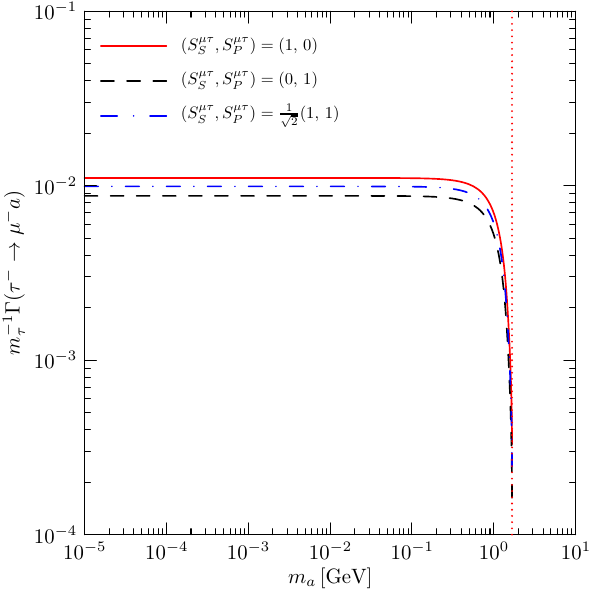}
}
\caption{
Fig.~\ref{fig:Axion:eFrac:pXi} and Fig.~\ref{fig:Axion:eFrac:nXi} show the
energy fraction distributions of electron and muon in decays 
$\tau^{-} \to e^{-} a $ with $\xi_{e\tau}^{+}  = \pm1$, respectively. 
The ALP has been assumed to be massless, and the $\tau$-leptons
are assumed to be generated with a fixed energy $E_{\tau} = 5.29\gev$.
Our predictions are shown by the curves, and the filled squares and circles
are simulation results using MadGraph with $3\times10^{4}$ events in total.
Fig.~\ref{fig:Axion:Gamma}, mass dependence 
of the decay width $\Gamma(\tau^{-}\to\mu^{-} a)$ for three configurations of the ALP
coupling constants: $(S_{S}^{\mu\tau},\, S_{P}^{\mu\tau}) = (1,\,0),\,(0,\,1), (1,\,1)/\sqrt{2}$.
}
\end{center}
\end{figure} 
which is simply a leaner function the energy fraction $z_{\ell}$, 
and depends on sign of $\xi_{\ell\ell'}^{+}$.
For positively polarized $\ell'$-lepton and $\xi_{\ell\ell'}^{+} > 0$, 
the decay events are dominated in the region with $\cos\theta_{\ell} < 0$, 
and hence have small energy fractions. In contrast, 
most of the $\ell$-leptons have larger energy fractions
if the $\ell'$-lepton is negatively polarized. 
Those polarization effects are shown 
in Fig.~\ref{fig:Axion:eFrac:pXi} and Fig.~\ref{fig:Axion:eFrac:nXi} 
for $\tau^{-} \to e^{-} a $ with $\xi_{\ell\ell'}^{+}  = \pm1$, respectively.
In both case, the ALP has been assumed to be massless, and
the $\tau$-lepton is boosted to have an energy $E_{\tau} = 5.29\gev$ which
corresponds to the case of $\tau$-lepton production in pair at the Belle II.
In our analytical calculations, mass of the $\ell$-lepton has been neglected,
so we can also see some differences near boundary between the numerical 
simulation results and our analytical predictions.
One more important property that we can learn form 
\eqref{eq:Axion:invDW:eFrac} is that, 
for relatively heavy ALP the $\gamma$-factor of $\ell$-lepton can be moderate,
then there is a cancellation between the first two terms in \eqref{eq:Axion:invDW:eFrac}
when the ALP interactions are dominated by the pseudo-scalar coupling
($S_{S}^{\ell\ell'} \ll S_{P}^{\ell\ell'}$). 
In this case the decay width is rapidly decreased and hence
giving smaller branching ratio. However, this happens in very narrow region because of
the large mass hierarchy in the lepton sector. For instance the decay process 
$\tau^{-} \to \mu^{-} a$, we can see in Fig.~\ref{fig:Axion:Gamma}, 
the three configurations of $(S_{S}^{\mu\tau},\, S_{P}^{\mu\tau})$ 
are nearly impossible to distinguish near the boundary $m_{a} = m_{\tau} - m_{\mu}$.

\subsection{Polarization Effects in Decays $\ell'^{\pm} \to \ell^{\mp} \imath^{\pm} \jmath^{\mp}$ }
\label{sec:AxionDecay}
In this subsection we consider the case of that the ALP undergoes further decays. 
Within our scenario (only flavor changing couplings are non-zero), 
the decaying particle has to be the $\tau$-lepton (\ie, $\ell'=\tau$), 
and for the decay products $\imath = e$ and $\jmath = \mu$. 
So the only possible channel is $\tau^{\pm} \to \ell^{\pm} (a^{\ast} \to \mu^{\pm} e^{\mp})$
with $\ell^{\pm} = \mu^\pm, e^\pm$.
Here we give only results of the decay process 
$\tau'^{-} \to \ell^{-} a^\ast \to \ell^{-} (e^{+} \mu^{-})$. 
Results of its charge conjugated processes can be obtained in a similar way.
Since the ALP is a scalar, its decay distribution in its rest frame is completely isotropic. 
Therefore, the decay density matrix elements can be written as,
\bee\label{eq:Axion:Decay:3b}
\calk_{ \lambda_{\ell'} }^{ \lambda^{'}_{\ell'} } 
=
\int_{ (m_{\mu} + m_{e})^{2} }^{ (m_{\tau} - m_{\ell})^{2} } \frac{ d m_{X}^{2} }{2\pi}
\frac{ 2 m_{X} \, \Gamma_{a\to e^{+} \mu^{-}}( m_{X}^{2} ) 
}{ (m_{X}^{2} - m_{a}^{2})^{2} + m_{a}^{2}\Gamma_{a}^{2} }\,
\cald_{ \lambda_{\ell'} }^{ \lambda^{'}_{\ell'} } \,,
\ene
where $m_{X}^{2} = (p_{e^+} + p_{\mu^-})^2$, $\Gamma_{a}$ is the
total decay width of $a$, and
$\cald_{ \lambda_{\ell'} }^{ \lambda^{'}_{\ell'} }$ are the helicity density 
matrix elements of the process $\ell'^{-} \to \ell^{-} a$ which have been given 
in last section. The decay width
$ \Gamma_{a\to e^{+} \mu^{-}}( m_{X}^{2} ) $ is simply given as,
\bee
\frac{1}{m_{X}}\Gamma_{a\to e^{+} \mu^{-}}( m_{X}^{2} )
=
\frac{ \overline{\beta_{a}} }{ 8 \pi  } 
\left( \epsilon_{+} \big|S_{S}^{\mu e}\big|^{2} + \epsilon_{-} \big|S_{P}^{\mu e}\big|^{2} \right)\,,
\ene
with $\epsilon_{\pm} = 1 - (m_{\mu} \pm m_{e})^{2}/m_{X}^{2} $ and
$\overline{\beta_{a}} = \sqrt{ 1 + (m_{\mu}^2 - m_{e}^2)^{2}/m_{X}^{4} - 2m_{e}^{2}/m_{X}^{2} 
- 2m_{\mu}^{2}/m_{X}^{2} }$. 
Because of the integrand in \eqref{eq:Axion:Decay:3b} is always positive, 
it is clear that polarization effects of the decay process
$\tau'^{-} \to \ell^{-} a^\ast \to \ell^{-} (e^{+} \mu^{-})$
is the same as the process $\tau^{-} \to \ell^{-} a $. This is also true even when
the decay $a\to e^{+} \mu^{-}$ is measured inclusively (this can naturally happens 
in practice when decay width of the ALP is sufficiently small, 
and hence become invisible inside of the 
detector~\cite{Heeck:2017xmg,Bjorkeroth:2018dzu}).
Within our conventions $S^{e\mu}_{S} = (S^{\mu e}_{S})^{\ast}$ and 
$S^{e\mu}_{P} = - (S^{\mu e}_{P})^{\ast}$, hence 
the total decay width $\Gamma_{a}=2 \Gamma_{a\to \mu^{+} e^{-}}( m_{a}^{2} )$.

If mass of the ALP is smaller then the $\tau$-lepton, 
then the dominate contribution is given when $m_{X}^{2} = m_{a}^2$. 
Because $m_\mu \gg m_e$, approximations of the kinematical factors
$\epsilon_{\pm}\approx \epsilon = 1 - m_{\mu}^{2}/m_{a}^{2}$ and 
$\overline{\beta_{a}} \approx  1 - m_{\mu}^{2}/m_{a}^{2}  = \epsilon$
should work well. In this limit, the decay width is simply given as,
$m_{X}^{-1}\Gamma_{a\to e^{+} \mu^{-}}( m_{a}^{2} ) =(8 \pi)^{-1} \epsilon^{2} S_{\mu e}^{+} $,
which depends only on sum of the squared ALP coupling constants.
For a typical ALP couplings at $\tev$ scale,
$m_{a}^{-1}\Gamma_{a\to\mu^{+} e^{-}}( m_{a}^{2} ) \le 4\times 10^{-2} S_{e\mu} \sim 10^{-10}$, which is rather small.
The total differential decay width is given as,
\bee
\frac{1}{ m_{\tau} }\frac{ d \Gamma_{ \lambda_{\tau}  } }{ d \cos\theta_{\ell}  }
=
\frac{ \overline{\beta}_{\tau\ell a}  }{ 512 \pi^{2} } 
\alpha_{a\ell}^{+} \alpha_{a\ell}^{-} \,\epsilon^{2}(m_{a}^{2})   S_{\ell \tau}S_{\mu e}^{+}
\left[
1 + \gamma_{\ell}^{-1} \kappa_{\ell \tau}^{-}  
-
\lambda_{\tau } \xi_{\ell\tau}^{+} \beta_{\ell} \cos\theta_{\ell}  \right] \,.
\ene
All polarization effects discussed in last section survive completely. 
In case of that $m_a > m_\tau$, proper integration with the variable $m_{X}^2$ 
is necessary. However, angular dependence of the total decay width does not change.

\subsection{Polarization Effects in Decays $\ell' \to \ell \bar{\nu}_{\ell} \nu_{\ell'}$ }
\label{sec:TauLepDecay}
\begin{figure}[h]
\begin{center}
\includegraphics[scale=0.682]{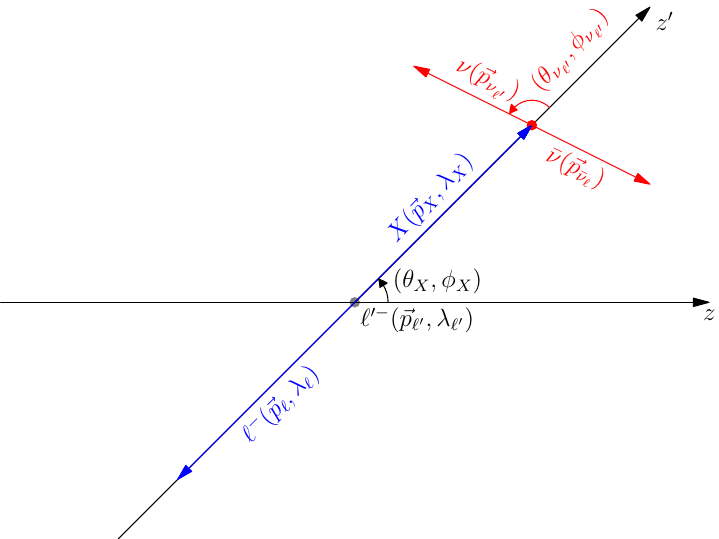}
\caption{
Definitions of the kinematical variables, 
particularly $p_{X}=p_{\nu_{\ell'}} + p_{\bar{\nu}_{\ell}}$, and $m_{X}^{2}=p_{X}^{2}$. 
The angular variables $(\theta_{X}, \phi_{X} )$ and $(\theta_{\nu_{\ell'}},\, \phi_{\nu_{\ell'}})$ 
of the momentum $p_{X}$ and $p_{\nu_{\ell'}}$ are defined in the rest frames of $p_{\ell'}$ 
and $p_{X}$, respectively. 
}
\label{fig:Neu:Inclusive:Kin}
\end{center}
\end{figure}

The leptonic decay modes $\ell' \to \ell \bar{\nu}_{\ell} \nu_{\ell'}$ 
are usually irreducible backgrounds of new physics signals.
For instance, both $\tau\to\mu\gamma$ and $\tau\to\ell\alpha$ suffer from 
heavy contamination of the SM process 
$\tau \to \ell \nu_{\tau} \bar{\nu}_{\ell}$~\cite{Villanueva:2018pbk,Konno:2020+A,Kou:2018nap}. 
However it is very hard to probe the signals because of more than one neutrino appear 
in the final state of the background. 
In this subsection, we study the full decay density matrix of the decay 
$\ell' \to \ell \bar{\nu}_{\ell} \nu_{\ell'}$ in
the rest frame of the decaying $\ell'$-lepton. On account of the missing neutrinos, 
this frame can not be accurately reconstructed in real experiments.
However, we will show that even when the relative momentum between the two neutrinos 
are integrated out, both longitudinal and transverse polarization effects survives. 
This is particularly important to the $\tau$-lepton, because of
its large leptonic decay branching ratio. Without loss of generality, all the formula 
are given for decay of $\ell'^{-}$-lepton, relevant results for decay of its anti-particle
can be obtained in a similar way. Furthermore, our analytical results will be 
given in mass less limit of the outgoing leptons and neutrinos. 
The kinematical variables are defined as follows (see Fig.~\ref{fig:Neu:Inclusive:Kin} for detials),
\bee
\ell'^{-}( \vec{p}_{\ell'}, \lambda_{\ell'}  ) 
\longrightarrow 
\nu_{\ell'}(\vec{p}_{\nu_{\ell'}} )
+
\bar{\nu}_{\ell}(\vec{p}_{\bar{\nu}_{\ell}})
+
\ell^{-}(\vec{p}_{\ell}, \lambda_{\ell} ) \,,
\ene
helicities of the $\ell'$-lepton 
and $\ell$-lepton take values of $\lambda_{\ell'}/2 = \pm1/2$ and 
$\lambda_{\ell}/2 = \pm1/2$, respectively.  

For the above leptonic decay, 
different from the usual phase space decomposition, 
here momentum of the two neutrino will be combined by introducing a virtual momenta 
$p^{\mu}_{X} = p^{\mu}_{\nu_{\ell'}} + p^{\mu}_{\bar{\nu}_{\ell} }$ 
(see Fig.~\ref{fig:Neu:Inclusive:Kin} for definitions of the kinematical variables,
and the explicit momentum parameterizations can be found in Sec.~\ref{app:kin:bkg}),
and the phase space factors are given as,
\bea
d\varPhi_{\ell'} 
&=& 
\frac{1}{2\pi} d m_{X}^{2} 
d\varPi_{\ell'}( \theta_{X}, \phi_{X} ) 
d\varPi_{X}(\theta_{\nu_{\ell'}},\, \phi_{\nu_{\ell'}}) \,,
\\[3mm]
d\varPi_{\ell'}
&=&
\frac{ d^{3} \vec{p}_{\ell}  }{ (2\pi)^{3} 2E_{\ell} }
\frac{ d^{3} \vec{p}_{X}  }{ (2\pi)^{3} 2E_{X} }
(2\pi)^{4}\delta^{4}( p_{\ell'} - p_{\ell} - p_{X} ) 
=
\frac{ \bar{\beta}_{X} }{8\pi}
\frac{ d\cos\theta_{X} }{2}
\frac{ d \phi_{X} }{ 2\pi } \,,
\\[3mm]
d\varPi_{X}
&=&
\frac{ d^{3} \vec{p}_{\nu_{\ell'} } }{ (2\pi)^{3} 2E_{\nu_{\ell'}} }
\frac{ d^{3} \vec{p}_{\bar{\nu}_{\ell} }}{ (2\pi)^{3} 2E_{\bar{\nu}_{\ell}} }
(2\pi)^{4}\delta^{4}( p_{X} - p_{\nu_{\ell'}} - p_{\bar{\nu}_{\ell} }) 
=
\frac{ 1 }{8\pi}
\frac{ d\cos\theta_{\nu_{\ell'}} }{2}
\frac{ d \phi_{\nu_{\ell'}} }{ 2\pi } \,.
\ena
Here the angular variables $\theta_{X} $ and $\phi_{X} $ are defined
in the rest frame of the parent $\ell'$-lepton, 
and the angular variables $\theta_{\nu_{\ell'}}$ and $\phi_{\nu_{\ell'}}$ are defined 
in the rest frame of the momentum $p_{X}^{\mu}$.

The decay density matrix in helicity basis can be written as,
\bee
\rho_{\lambda_{\ell'}}^{\lambda'_{\ell'}}
= 
\calm_{ \lambda_{\ell'} } \calm^{\dag}_{ \lambda'_{\ell'} } \,,
\ene
where $\calm_{ \lambda_{\ell'} }$ is the decay helicity amplitude. 
Assuming that decay of the $\ell'$-lepton is completely described by the standard model (SM),
then the decay helicity amplitude is given by contract of the charged currents of the
$\ell'-\nu_{\ell'}$ and $\ell-\nu_{\ell}$ systems. In this form, the two neutrinos
couple separately to the two leptons, and hence it is relatively hard to obtain the
inclusive decay density matrix. In order to integrate out the relative
kinematical variables of these two neutrinos, we apply Fierz transformation 
for the decay helicity amplitude which in turn is given by a contract 
of two neutral currents of the $\ell'-\ell$ and $\nu_{\ell'}-\nu_{\ell}$ systems,
\bea
\calm_{ \lambda_{\ell'} }
&=& 
\frac{ g_{W}^{2} }{2 D_{W} }
\caln^{\mu}_{ \lambda_{\ell'} }(\ell', \ell )  \,
\caln_{\mu}(\nu_{\ell'}, \bar{\nu}_{\ell})  \,,
\\[3mm]
\caln^{\mu}_{ \lambda_{\ell'} }(\ell', \ell ) 
&=&
\overline{ u_{\ell}(\vec{p}_{\ell} ) }  
\gamma^{\mu}\gamma_{L} 
u_{\ell'}(\vec{p}_{\ell'}, \lambda_{\ell'}) \,,
\\[3mm]
\caln^{\mu}(\nu_{\ell'}, \bar{\nu}_{\ell})
&=&
\overline{ u_{\nu_{\ell'}}(\vec{p}_{\nu_{\ell'}} ) }  
\gamma^{\mu}\gamma_{L} 
v_{\bar{\nu}_{\ell}}(\vec{p}_{\bar{\nu}_{\ell}}) \,.
\ena
where $D_{W}^{-1}$ is propagator of the charged weak bosons.
Accordingly, the decay spin density matrix can be rewritten as,
\bee\label{eq:pure:density}
\rho_{\lambda_{\ell'}}^{\lambda'_{\ell'}} 
= 
\frac{ g_{W}^{4} }{4 D_{W}^{2} }  
\caln^{\lambda'_{\ell'} \mu\nu}_{\lambda_{\ell'}} (\ell', \ell ) 
\caln_{\mu\nu}(\nu_{\ell'}, \bar{\nu}_{\ell}) \,,
\ene
where the two tensors are defined as,
\bea
\caln^{\lambda'_{\ell'} \mu\nu}_{\lambda_{\ell'}} (\ell', \ell )
&=& 
\caln^{\mu}_{ \lambda_{\ell'} }(\ell', \ell ) \, 
\caln^{\nu\dag}_{ \lambda'_{\ell'} }(\ell', \ell ) \,,
\\[3mm]
\caln_{\mu\nu}(\nu_{\ell'}, \bar{\nu}_{\ell})
&=&
\caln_{\mu}(\nu_{\ell'}, \bar{\nu}_{\ell}) \,  \caln_{\nu}^{\dag}(\nu_{\ell'}, \bar{\nu}_{\ell}) \,.
\ena
In this expression, relative momentum of the two neutrinos can be formally integrated out. 
In general the factor $D_{W}^{-2}$ in \eqref{eq:pure:density}
can also depend on the angular variable $\theta_{\nu_{\ell'}}$. 
Then the relative momentum of the two neutrinos can be integrated out by defining
\bee
\overline{\caln}^{\mu\nu}(m_{X}^{2})
=
\frac{1}{4} g_{W}^{4} \int d\varPi_{X} \, D_{W}^{-2}\; 
\caln^{\mu\nu}(\nu_{\ell'}, \bar{\nu}_{\ell}) \,,
\ene
which is just a function of the invariant mass $m_{X}^{2} $. Correspondingly, 
the inclusive decay spin density matrix is given as,
\bee
\overline{\rho}_{\lambda_{\ell'}}^{\lambda'_{\ell'}} 
= 
\caln^{\lambda'_{\ell'} \mu\nu}_{\lambda_{\ell'}} (\ell', \ell ) 
\, \overline{\caln}^{\mu\nu}(m_{X}^{2} ) .
\ene
In this sense, the tensor $\overline{\caln}^{\mu\nu}(m_{X}^{2})$ is the most
essential quantity for studying spin correlation effects in the decay process.
The tensor $\overline{\caln}^{\mu\nu}(m_{X}^{2})$ in the rest frame of the $\ell'$-lepton
can be obtained by calculating its expression in the rest frame of the momentum $p_{X}$,
and then applying rotations and boosts. With the excellent approximation,
$
D_{W} 
\approx - m_{W}^{2} + i m_{W}\varGamma_{W}
$,
the angular variables $(\theta_{\nu_{\ell'}},\, \phi_{\nu_{\ell'}})$ can be integrated out directly. 
Our calculation gives,
\bee\label{eq:inc:nut:ten}
\overline{\caln}^{\mu\nu}( m^{2}_{X} )  
=
f_{\ell}
\left(
p^{\mu}_{X} p^{\nu}_{X} -  m_{X}^{2} g^{\mu\nu}
\right) \,,
\ene
where the coefficient
$f_{\ell} = g_{W}^{4} [ 96\pi m^{2}_{W}(m_{W}^{2} +  \varGamma^{2}_{W} ) ]^{-1} $

\begin{figure}[ht]
\begin{center}
\subfigure[]{
\label{fig:Neu:Inclusive:Polarizer}
\includegraphics[width=0.312\linewidth]{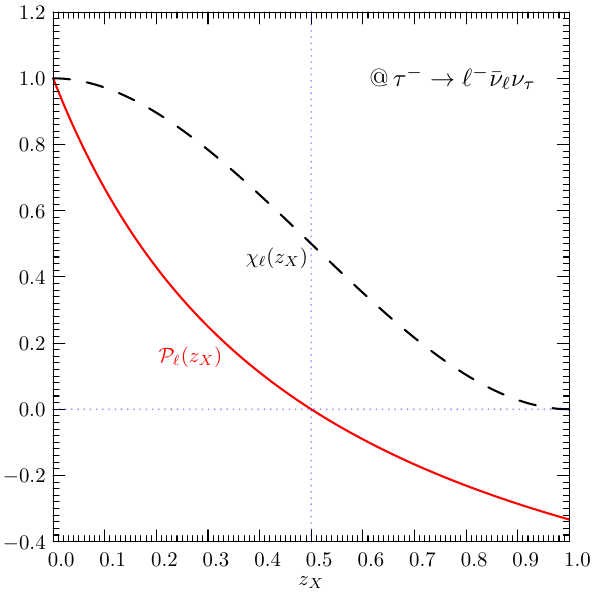}
}
\subfigure[]{
\label{fig:Neu:Inclusive:massX}
\includegraphics[width=0.312\linewidth]{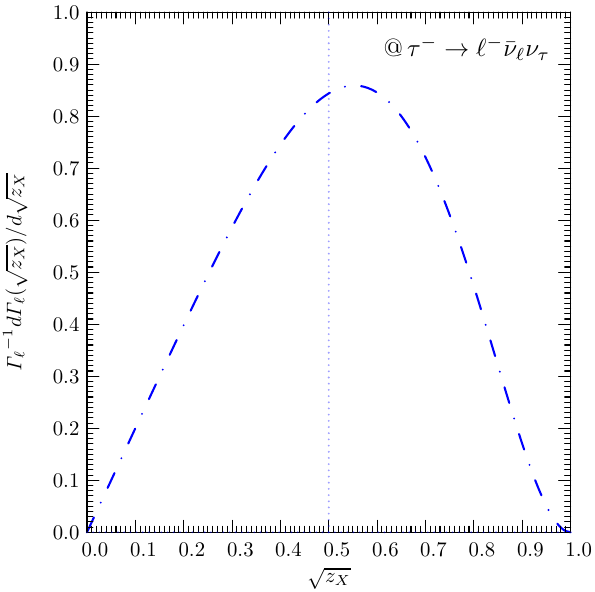}
}
\subfigure[]{
\label{fig:Axion:eFracTauToMuo}
\includegraphics[width=0.312\linewidth]{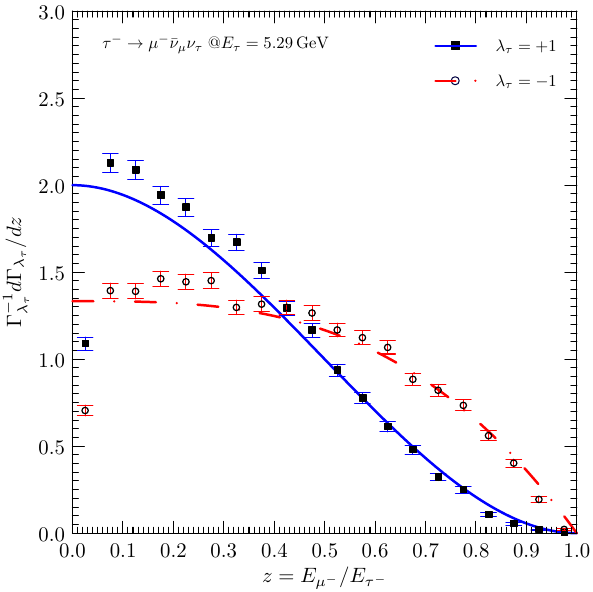}
}
\caption{Fig.~\ref{fig:Neu:Inclusive:Polarizer}, plots of the functions 
$\chi( z_{X} )$ and $\calp_{\ell^{-}}( z_{X} )$. 
Fig.~\ref{fig:Neu:Inclusive:massX}, differential decay width with respect to the 
normalized invariant mass, $\sqrt{z_{X}}=m_{X}/m_{\tau}$, of the two neutrino system. 
Fig.~\ref{fig:Axion:eFracTauToMuo}, energy fraction distributions of $\mu^{-}$  
in decay $\tau^{-} \to \mu^{-} \bar{\nu}_{\mu} \nu_{\tau}$ 
with different polarizations $\lambda_{\tau}=\pm1$ of the $\tau$-leptons. 
The $\tau$-leptons are assumed to be generated with a fixed energy 
$E_{\tau} = 5.29\gev$. Our predictions given by \eqref{eq:TauLepDec:eFrac} 
are shown by the curves. The filled squares and circles
are simulation results by using MadGraph with 
$3\times10^{4}$ events in total.
}
\end{center}
\end{figure}

The neutral current $\caln^{\mu}_{ \lambda_{\ell'} }(\ell', \ell )$ 
can be calculated directly in the rest frame of the $\ell'$-lepton, and is given as,
\bee\label{hel:tau:vector}
\caln^{\mu}_{ \lambda_{\ell'} }
=
\sqrt{ \overline{\beta}_{X} } \; 
\left(
\begin{array}{c}
e^{\frac{i\lambda_{\ell'}\phi_{X}}{2}} 
\sqrt{ \dfrac{ 1+\lambda_{\ell'}\cos\theta_{X} } { 2\overline{\beta}^{-1}_{X}} }
\\[6mm]
 - e^{\frac{-i\lambda_{\ell'}\phi_{X}}{2}}  
 \sqrt{ \dfrac{ 1-\lambda_{\ell'}\cos\theta_{X}}{2\overline{\beta}^{-1}_{X}}} 
\\[6mm]
-i\lambda_{\ell'} e^{\frac{-i\lambda_{\ell'}\phi_{X}}{2}}  
\sqrt{ \dfrac{ 1-\lambda_{\ell'}\cos\theta_{X}} {2\overline{\beta}^{-1}_{X}} } 
\\[6mm]
-\lambda_{\ell'} e^{\frac{-i\lambda_{\ell'}\phi_{X}}{2}} 
\sqrt{ \dfrac{ 1+\lambda_{\ell'}\cos\theta_{X}} {2\overline{\beta}^{-1}_{X}} }
\end{array}
\right)\,.
\ene
By using following relations,
\bes\label{eq:dens:con}
\bea 
p_{X} \cdot \caln_{ \lambda_{\ell'} }
&=&
\dfrac{ m_{\ell'}^{2} }{ \sqrt{2} }
\; e^{ \frac{ i\lambda_{\ell'}\phi_{X} }{2}}
\sqrt{ \dfrac{ 1+\lambda_{\ell'}\cos\theta_{X} }{ \overline{\beta}_{X}^{-1} } }  \,,
\\[3mm]
\caln_{ \lambda_{\ell'} } \cdot \caln^{\dag}_{ \lambda_{\ell'} }
&=&
- m_{\ell'}^{2} \overline{\beta}_{X}\big(1-\lambda_{\ell'}\cos\theta_{X} \big)  \,,
\\[3mm]
\caln_{ \lambda_{\ell'} } \cdot \caln^{\dag}_{ \lambda_{\ell'} }
&=&
m_{\ell'}^{2} \overline{\beta}_{X} \sin\theta_{X}  e^{i\lambda_{\ell'}\phi_{X}} \,.
\ena
\ens
Then we can obtain the density matrix,
\bea
\overline{\rho}_{\lambda_{\ell'}}^{\lambda_{\ell'}} 
&=& 
f_{\ell} m_{\ell'}^{4} \overline{\beta}_{X}^{-1} 
\chi( z_{X}  )\bigg(1+\lambda_{\ell'} \calp_{\ell}( z_{X} ) \cos\theta_{X}\bigg)\,,
\\[3mm]
\label{eq:dens:off}
\overline{\rho}_{\lambda_{\ell'}}^{-\lambda_{\ell'}} 
&=&  
f_{\ell}  m_{\ell'}^{4} \overline{\beta}_{X}^{-1} \chi( z_{X}  )
\calp_{\ell}( z_{X} ) \sin\theta_{X}  e^{i\lambda_{\ell'}\phi_{X}} \,,
\ena
where $z_{X} = m_{X}^{2}/m_{\ell'}^{2}$, and the coefficient are given as,
\bea
\chi( z_{X} )  
&=&
\big(  1 -  z_{X}   \big)^{2} 
\big(  1 + 2 z_{X}   \big)  \,,
\\[3mm]
\calp_{\ell}( z_{X} ) 
&=&
\frac{ 1 - 2 z_{X}  }{ 1 + 2 z_{X}  }  \,.
\ena
We can see that both longitudinal and transverse spin correlations survive when
the relative momentum was integrated out. The transverse correlation is powerful
for probing CP violation effect. We will investigate elsewhere its
possible application in searching $CP$-violation effect in decay of the Higgs boson 
$h(125)$~\cite{Aad:2012tfa,Chatrchyan:2012xdj} to $\tau$-lepton pair~\cite{Hagiwara:2016zqz}.
Here we study only the longitudinal polarization effect.
Fig.~\ref{fig:Neu:Inclusive:Polarizer} shows the functions $\chi( z_{X} )$ and 
$\calp_{\ell^{-}}( z_{X} )$. 
We can see that the polarizor $\calp_{\ell^{-}}( z_{X} )$ decrease quickly 
with respect to $z_{X}$, and change its sign at  $z_{X} \approx 0.5$. 
This is due to that contributions of the spin-0 and the longitudinal component of the spin-1 
states have different sign (see Eq.~\eqref{eq:inc:nut:ten}), and hence the $m_{X}$-dependent
and $p_{X}$-dependent terms interfere deconstructively (see Eq.~\eqref{eq:dens:con}).
Fig.~\ref{fig:Neu:Inclusive:massX} shows the normalized differential decay width 
$\varGamma_{\ell}^{-1}d\varGamma_{\ell}/d\sqrt{z_{X}}$. 
Firstly, we can see that decay rate with harder invariant mass $m_{X}$
has a relatively larger fraction. 
Since correlation between $m_{X}$ and the energy of the lepton, 
leptons are expected to have a relatively soft energy spectrum in the laboratory frame. 
This property can reduce experimental efficiency due to cuts on momentum.
However it is not too worse compared to the decay mode $\tau \to \pi\nu_{\tau}$ 
since usually harder cuts are used for the hadronic decay mode.

The 2-D differential decay width is given as,
\bee
\frac{ d \varGamma_{\lambda_{\ell'}} }{ d z_{X} d\cos\theta_{\ell} }
=
\frac{ G_{F}^{2} m_{\ell' }^{5}    }{192\pi^{3} } 
\chi( z_{X}  )\bigg(1 - \lambda_{\ell'} \calp_{\ell}( z_{X} ) \cos\theta_{\ell}\bigg)\,,
\ene
where $G_{F} = g_{W}^{2}/(4\sqrt{2}m_{W}^{2})$ is the Fermi constant 
(we have used approximation $\Gamma_{W} \approx 0$), 
and we have used the relations $\theta_{X} = \pi -\theta_{\ell}$ such that the distributions
are expressed in term of the polar angle of the charged lepton.
Integrating over the invariant mass $m_{X}$, 
the averaged polarizer $\overline{\calp}_{\ell}=1/3$. 
Compared to the hadronic decay mode 
$\tau\to\pi\nu_{\tau}$, $\overline{\calp}_{\ell}$ is smaller. Hence the leptonic
decay mode has lower sensitivity to polarization of the parent $\tau$-lepton. 
However, in practice the leptonic decay mode can also be important
because of its much larger branching ratio.
In term of energy fraction $z_{\ell}$, the 2-D differential decay width is given 
as,
\bee
\frac{ d \varGamma_{\lambda_{\ell'}} }{ d z_{X} dz_{\ell} }
=
\frac{ G_{F}^{2} m_{\ell' }^{5}    }{48\pi^{3} } 
\beta_{\ell'}^{-1} \overline{\beta}_{X}^{-1}
\chi( z_{X}  )\left[ 
\frac{1}{2} \left( 1 + \lambda_{\ell'} \beta_{\ell'}^{-1} \beta_{\ell}^{-1}\calp_{\ell}( z_{X} )  \right) 
- \lambda_{\ell'} \beta_{\ell'}^{-1} \overline{\beta}_{X}^{-1}\calp_{\ell}( z_{X} ) z_{\ell}\right]\,.
\ene
Here we should note that upper and lower integration limits of the variable $z_{\ell}$ depend
on $z_{X}$, and are given as ${\rm Max} \{z_{\ell} \} = \alpha_{\ell} ( 1 + \beta_{\ell'}\beta_{\ell}  )/2$
and ${\rm Min} \{z_{\ell} \} = \alpha_{\ell} ( 1 - \beta_{\ell'}\beta_{\ell}  )/2$. 
In the limit $\beta_{\ell'} \approx 1$, ${\rm Max} \{z_{\ell} \} \approx  1 + z_{X}$ and
${\rm Min} \{z_{\ell} \} \approx  0$. Within this limit,
and after integration with respect to the variable $z_{X} \in [0,\,1 - z_{\ell} ]$ we obtain,
\bee\label{eq:TauLepDec:eFrac}
\frac{ d \varGamma_{\lambda_{\ell'}} }{ dz_{\ell} }
=
\frac{ G_{F}^{2} m_{\ell' }^{5}    }{192\pi^{3} } 
\left[ \cali(z_{\ell}) + \lambda_{\ell'} \calk(z_{\ell}) \right]\,,
\ene
where the two functions are given as
\bea
\cali(z_{\ell}) &=& \frac{1}{3} \left(  4 z_{\ell}^3 - 9 z_{\ell}^2 + 5 \right) \,,
\\[3mm]
\calk(z_{\ell})  &=& \frac{1}{3} \left(  8 z_{\ell}^3 - 9 z_{\ell}^2 + 1 \right) \,.
\ena
The above two functions have been normalized such that integration of 
$\cali(z_{\ell})$ with respect to $z_{\ell} \in [0,\, 1]$ gives 1, 
and integration of $\calk(z_{\ell})$ in the same integration is zero as it should be. 
Fig.~\ref{fig:Axion:eFracTauToMuo} shows
energy fraction distribution of the $\mu^{-}$  in decay
$\tau^{-} \to \mu^{-} \bar{\nu}_{\mu} \nu_{\tau}$ 
with different polarizations $\lambda_{\tau} = \pm1$ of the $\tau$-leptons. 
The $\tau$-leptons are assumed to be generated with a fixed energy 
$E_{\tau} = 5.29\gev$. We can see that after integration over the invariant mass
$m_{X}$, angular distributions for $\tau$-lepton with different polarizations
are still distinct, and also clearly different from distributions of the signal
as shown in Fig.~\ref{fig:Axion:eFrac:pXi} and Fig.~\ref{fig:Axion:eFrac:nXi}.

\section{Polarization Correlation Effects}
\label{sec:correlations}
In this section we study the spin correlation effects in decays of $\tau$-lepton pair 
which is produced at a $e^{-}e^{+}$ collider. 
Particularly we focus on the Belle II experiment 
where collision energy $\sqrt{s} = 10.58\gev$ in the center of mass frame.
At the Belle II, the $\tau$-lepton pair is produced essentially via a virtual photon, 
and hence is unpolarized. 
Nevertheless, polarizations of the two $\tau$-leptons are correlated because
helicities of the $\tau$-lepton pair are required to satisfy the relation 
$\lambda_{\tau} = - \lambda_{\overline{\tau}}$ 
thanks to spin conservation (in massless limit). 
As we have shown that polarization effects of the signal and backgrounds 
have distinctive property,
therefore, polarization correlations among these two $\tau$-leptons
can be powerful probes for detecting signals.
In this work we investigate following processes:
\bee
e^{-}e^{+} \to \tau^{-}_{\alpha}   \tau^{+}_{\beta}\,,\; 
\ene
where $\alpha, \beta = \ell, \pi, a\ell$ with
$\tau^{\pm}_{\ell} = \tau^{\pm} \to \ell^{\pm} \bar{\nu}_{\ell}\nu_{\tau}$, 
$\tau^{\pm}_{\pi} =\tau^{\pm} \to \pi^{\pm} \nu_{\tau}$ and  
$ \tau^{\pm}_{a \ell} = \tau^{\pm} \to \ell^{\pm} a $.
The other hadronic decay channels of the $\tau$-lepton can also be employed 
in a similar way, but here we study only the above channels for illustrating
usefulness of the polarization correlation.
We will study experimental sensitivities of all the channels elsewhere.
In any case the normalized differential cross section can be written as,
\bee
\frac{1}{ \sigma_{ \tau_{\alpha}^{-} \tau_{\beta}^{+} } }
\frac{ d\sigma_{ \tau_{\alpha}^{-} \tau_{\beta}^{+} } }
{ d z_{\alpha}^{-} d z_{\beta}^{+} }
=
\cali_{\tau^{-}_{\alpha} } 
\cali_{\tau^{+}_{\beta} }
+ 
\calk_{\tau^{-}_{\alpha} }( z_{\alpha}^{-}   )  
\calk_{\tau^{+}_{\beta} }( z_{\beta}^{+}  )  \,.
\ene

\begin{figure}[bt]
\begin{center}
\subfigure[]{
\label{fig:LonCor:EleAle}
\includegraphics[width=0.312\linewidth]{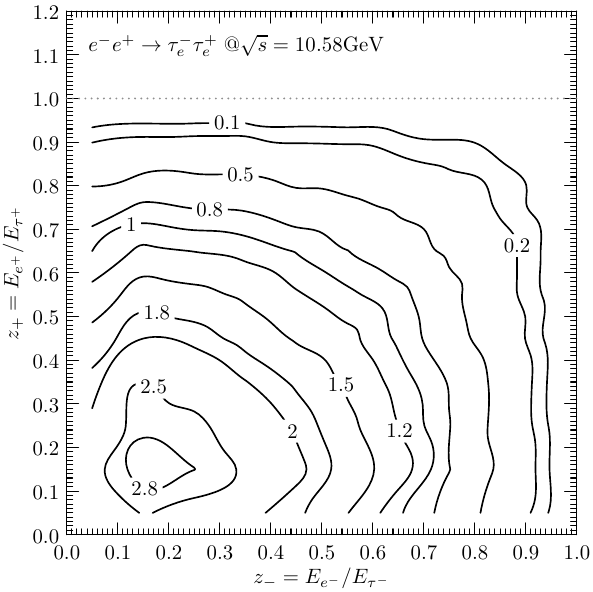}
}
\subfigure[]{
\label{fig:LonCor:EleXlePos}
\includegraphics[width=0.312\linewidth]{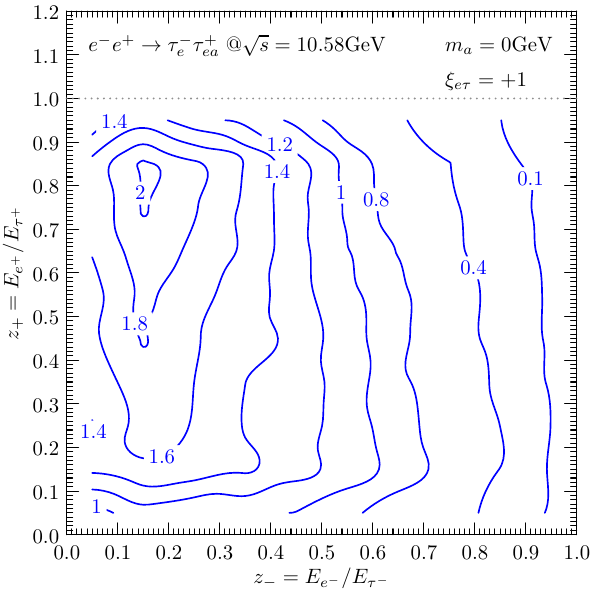}
}
\subfigure[]{
\label{fig:LonCor:EleXleNeg}
\includegraphics[width=0.312\linewidth]{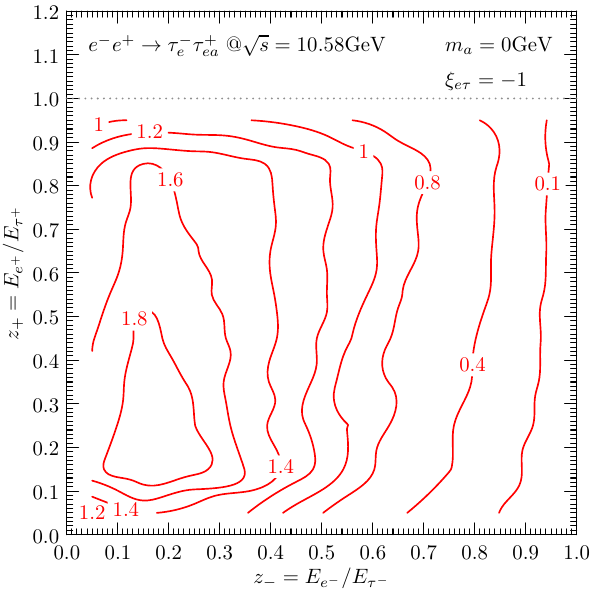}
}
\caption{
Contour plots of the normalized differential cross sections in the $z_{-} - z_{+}$ plane 
for the channels (a) $e^{-}e^{+} \to \tau^{-}_{e}   \tau^{+}_{e}$,
(b) and (c) for $e^{-}e^{+} \to \tau^{-}_{e}   \tau^{+}_{ae}$ with  $\xi_{e\tau} = \pm1$,
and $m_{a} = 0\gev$ at $\sqrt{s} = 10.58\gev$. 
The distributions are obtained by using $10^{5}$ events for every channels.
}
\end{center}
\end{figure}

For a leptonic channel,
$e^{-}e^{+} \to \tau^{-}_{\ell}   \tau^{+}_{\jmath}$, 
we have $\cali_{\tau^{-}_{\ell}} = \cali_{\tau^{+}_{\jmath}}=\cali$ and
 $\calk_{\tau^{-}_{\ell}} = \calk_{\tau^{+}_{\jmath}}=\calk$
in massless limits of the leptons $\ell$ and $\jmath$. 
In this case, there is a relatively soft positive correlation between 
$z_{\ell}^{-} $ and $z_{\jmath}^{+}$. Fig.~\ref{fig:LonCor:EleAle} 
shows contour plot of the above differential cross section
in $z_{\ell}^{-} - z_{\jmath}^{+}$ plane for the channel
$e^{-}e^{+} \to \tau^{-}_{e}   \tau^{+}_{e}$. 
We can see clearly the positive correlation.
Similar correlation effects can be found when either one of the two electrons 
or both electrons are replaced
by $\mu$-leptons, except for a small division due to their mass difference.
For the corresponding signal processes, 
$e^{-}e^{+} \to \tau^{-}_{\ell}   \tau^{+}_{a\jmath}$ and
$e^{-}e^{+} \to \tau^{-}_{a\ell}   \tau^{+}_{\overline{\jmath}}$, 
the differential cross section has the same form, but the polarization function 
$\calk_{\tau^{\pm}_{a\ell} }$ depends on the parameters $\xi_{\ell\tau}^{+}$
and $m_a$. Without loss of generality we discuss only the channel 
$e^{-}e^{+} \to \tau^{-}_{\ell}   \tau^{+}_{a\jmath}$,
completely the same distributions can be found in the charge conjugated channels.
Fig.~\ref{fig:LonCor:EleXlePos} and Fig.~\ref{fig:LonCor:EleXleNeg} 
show contour plots of the normalized differential cross section for the
signal process $e^{-}e^{+} \to \tau^{-}_{e}   \tau^{+}_{ae}$ 
with $m_{a} = 0\gev$ and $\xi_{e\tau} = \pm1$, respectively. 
We can see that the correlation properties are completely different from 
the corresponding background.

\begin{figure}[ht]
\begin{center}
\subfigure[]{
\label{fig:LonCor:PioAuo}
\includegraphics[width=0.312\linewidth]{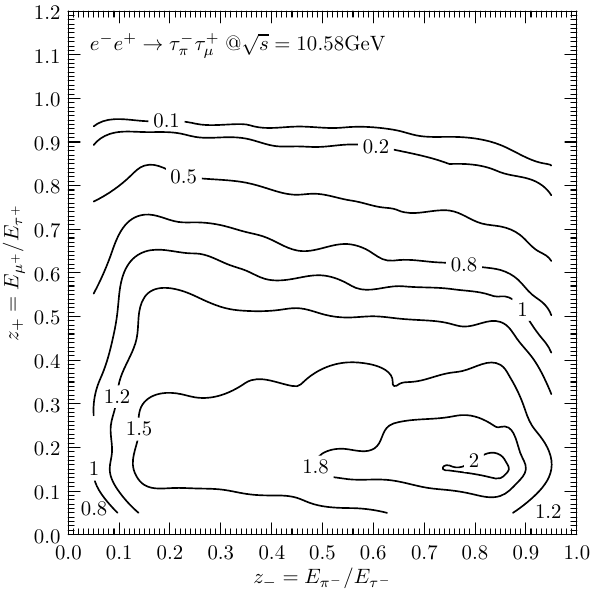}
}
\hfill
\subfigure[]{
\label{fig:LonCor:PioXuoPos}
\includegraphics[width=0.312\linewidth]{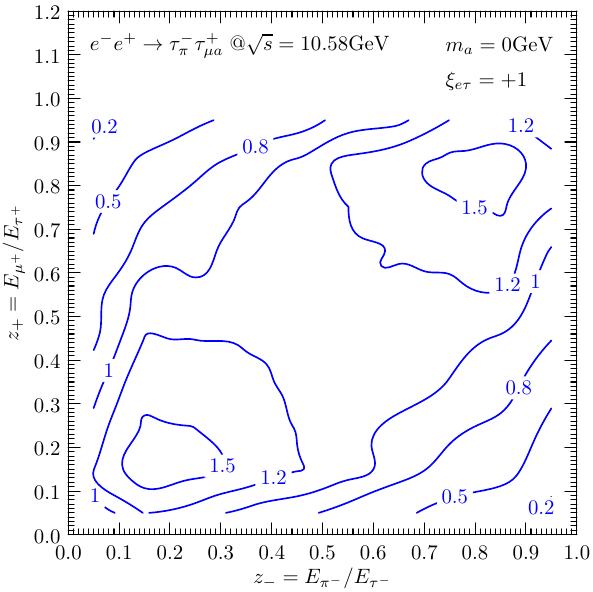}
}
\hfill
\subfigure[]{
\label{fig:LonCor:PioXuoNeg}
\includegraphics[width=0.312\linewidth]{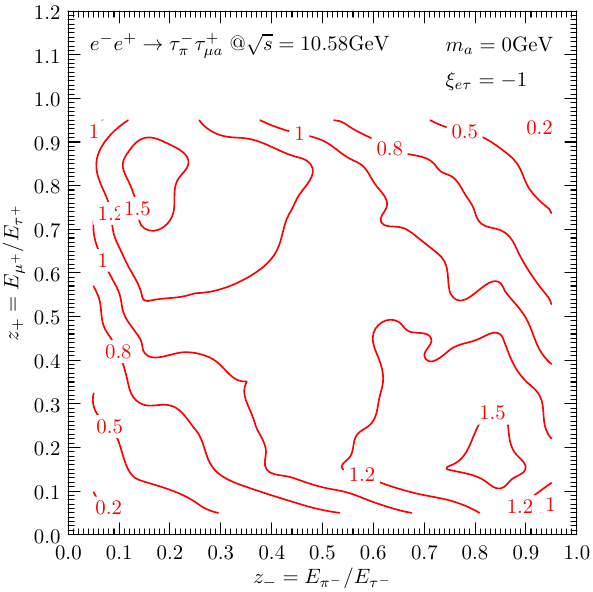}
}
\caption{
Contour plots of the normalized differential cross sections in the $z_{-} - z_{+}$ plane 
for the channels (a) $e^{-}e^{+} \to \tau^{-}_{\pi}   \tau^{+}_{\mu}$,
(b) $e^{-}e^{+} \to \tau^{-}_{\pi}   \tau^{+}_{a\mu}$  with $\xi_{\mu\tau} = 1$,
and (c) $e^{-}e^{+} \to \tau^{-}_{\pi}   \tau^{+}_{a\mu}$  with $\xi_{\mu\tau} = -1$
at $\sqrt{s} = 10.58\gev$. And we have used $m_{a} = 0\gev$ as an example.
The distributions are obtained by using $10^{5}$ events for every channels.
}
\end{center}
\end{figure}

Searches for the lepton flavor violation in decays of the $\tau$-lepton 
can be further improved by including hadronic decay modes of the $\tau$-lepton.
Fig.~\ref{fig:LonCor:PioAuo} shows polarization correlation in the process
$e^{-}e^{+} \to \tau^{-}_{\pi}   \tau^{+}_{\mu}$, whose correlation
is dominated by polarization effects of the decay mode $ \tau^{-}_{\pi}$.
In contrast, correlation of the signal process 
$e^{-}e^{+} \to \tau^{-}_{\pi}   \tau^{+}_{a\mu}$,
shows a simple positive or negative 
linear property, which are shown in Fig.~\ref{fig:LonCor:PioXuoPos} and Fig.~\ref{fig:LonCor:PioXuoNeg}
with $\xi_{\mu\tau} = \pm 1$, respectively.
The correlation behavior depends on the parameters $\xi_{\mu\tau}$
and $m_{a}$.
The model parameters can be extracted by using general fitting
method (for instance the likelihood method) in the $z_{-}-z_{+}$ plane. 
Comparing to the 3-prong tagging method used by 
the ARGUS collaboration~\cite{Albrecht:1995ht}, 
and the $3-1$ prong searching method used by the Belle II 
experiment~\cite{Abe:2010gxa,Villanueva:2018pbk,Konno:2020+A,Kou:2018nap},
as well as other double-side tagging proposals~\cite{DeLaCruz-Burelo:2020ozf,Xiang:2016jni,Christensen:2014yya}, 
our approach can significantly increase the available number of events.
We will study elsewhere the experimental sensitivities of all the channels
at the Belle II experiment.

\section{Summary}
\label{sec:conclusion}
In summary we have studied polarization effects of the lepton flavor violated decays 
induced by an ALP. Experimental searches for these rare decay channels are challenging 
because of the huge irreducible backgrounds $\ell' \to \ell \bar{\nu}_{\ell} \nu_{\ell'}$, 
and hence the current experimental bounds on the decay modes 
$\ell^{'\pm} \to a \ell^{\pm} $ are much weaker than other flavor violated processes.
In this paper, we treat inclusively the two missing neutrinos of the backgrounds,
and their relative momentum are integrated. After that we find that
both longitudinal and transverse polarization effects survives. 
Observing transverse polarization effects requires reconstructions 
of the mother leptons, and hence is challenging, particularly 
for $\tau$-leptons produced at low energy colliders, we will study this elsewhere. 
In contrast, energy fraction of its charged decay product
is correlated with the longitudinal polarization of the mother lepton. 
At $e^{-}e^{+}$ collider, it can be defined measured precisely,
and hence serves as an excellent observable of the polarization effect.
We show that signals and backgrounds have distinctive energy fraction distributions,
and hence can be employed to searching for lepton flavor violation.

On the other hand, leptons generated at low energy collider are usually unpolarized.
Hence there is no net polarization effect.
However, we show that polarization correlations between the $\tau$-leptons produced 
in pair are still useful to search for the signals. 
More interestingly, polarization correlations depends on product of 
scalar and pseudo-scalar ALP couplings, and hence are sensitive to their relative sign.
This is a distinctive property compared to 
the usual observable (cross section or decay branching ratio).
Moreover, because kinematical reconstruction as well as tagging of
the decaying lepton are not necessary,
number of the available signal events was increased significantly
in our approach.

\appendix

\section{Definitions related to the decay $\ell' \to \ell a$}
\label{app:kin:LTA}
For completeness and clarity, we give our parameterizations of 
the momentum and helicity amplitudes of the decay $\ell' \to \ell a$. 
In the rest frame of the $\ell'$-lepton, 
the momentum can be parameterized as follows,
\bea
p_{\ell'}^{\mu} 
&=&
\;\;\,m_{\ell'}\left(1,\,\;\;\;\,  0,\,\;\;\;\,  0,\,\;\;\;\,  0 \right)\,,
\\[3mm]
p_{a}^{\mu}
&=&
\frac{1}{2} m_{\ell'} \alpha_{a\ell}^{+} \left(
1, 
\;\;\;\, \beta_{a} \sin \theta_{a} \cos \phi_{a}, 
\,\;\;\,  \beta_{a} \sin \theta_{a} \sin \phi_{a}, 
\,\;\; \beta_{a}  \cos \theta_{a}\right)  \,,
\\[3mm]
p_{\ell^{-}}^{\mu}
&=&
\frac{1}{2} m_{\ell'}\alpha_{a\ell}^{-}  \left(
1 ,\, 
- \beta_{\ell} \sin \theta_{a} \cos \phi_{a}, \,
-  \beta_{\ell} \sin \theta_{a} \sin \phi_{a}, \,
-  \beta_{\ell} \cos \theta_{a}\right)  \,,
\ena
where $\alpha_{a\ell}^{\pm} = 1 \pm (m_{a}^{2} - m_{\ell}^{2})/m_{\ell'}^{2}$, 
and $\beta_{a} = \overline{\beta}_{\ell'\ell a}/\alpha_{a\ell}^{+}$ and
$\beta_{\ell} = \overline{\beta}_{\ell'\ell a}/\alpha_{a\ell}^{-}$ with 
$\overline{\beta_{\ell'\ell a}} = \sqrt{ 1 + (m_{\ell}^2 - m_{a}^2)^{2}/m_{\ell'}^{4} - 2m_{\ell}^{2}/m_{\ell'}^{2} - 2m_{a}^{2}/m_{\ell'}^{2} }$. 
According to the Lagrangian \eqref{eq:Model:Lep:Sim}, the helicity amplitudes are given as,
\bee
\cald_{\lambda_{\ell'}, \lambda_{\ell}}
=
\overline{ u(\vec{p}_{\ell}, \lambda_{\ell} ) }
\left( S_{S}^{\ell \ell'} + S_{P}^{\ell \ell'}\gamma_{5} \right) 
u( \vec{p}_{\ell'}, \lambda_{\ell}  ) \,.
\ene
With our parameterizations the momentum then we have,
\bea
\cald_{\lambda_{\ell'},+} 
&=&
\frac{ \lambda_{\ell'}  m_{\ell'} }{2} \sqrt{ \alpha_{a\ell}^{+} \alpha_{a\ell}^{-}  }
\left( f_{+}  S_{S}^{\ell \ell'} - f_{-}  S_{P}^{\ell \ell'}  \right) 
e^{-i\phi_{a}/2}e^{i\lambda_{\ell'}\phi_{a}/2} \sqrt{ 1 - \lambda_{\ell'} \cos\theta_{a} }\,,
\\[3mm]
\cald_{\lambda_{\ell'},-} 
&=&
\frac{  m_{\ell'}  }{ 2 } \sqrt{ \alpha_{a\ell}^{+} \alpha_{a\ell}^{-}  }
\left(   f{+} S_{S}^{\ell \ell'} +  f_{-} S_{P}^{\ell \ell'}  \right) 
e^{i\phi_{a}/2}e^{i\lambda_{\tau^{-}}\phi_{a}/2} \sqrt{ 1 + \lambda_{\ell'} \cos\theta_{a} } \,,
\ena
where $f_{\pm} = \left( \sqrt{ 1 +\beta_{\ell}  } \pm \sqrt{ 1 - \beta_{\ell}  } \right)/\sqrt{2}$, and
in massless limit of the lepton $\ell$ (which generally works well for studying polarization effect), 
$f^{\pm} = 1$. For completeness, we will keep the mass dependence in the density matrix elements.

\section{Definitions related to the decay $\ell' \to \ell \bar{\nu}_{\ell} \nu_{\ell'}$,}
\label{app:kin:bkg}
The related momentum in the rest frame of the $\ell'$-lepton are parameterized
as follows
\bea
p_{\ell'}
&=&
m_{\ell'} \left(1, 0, 0, 0  \right)  \,,
\\[3mm]
p_{X}^{\mu}
&=&
\frac{1}{2} m_{\ell'} \left(
\alpha_{X}, 
\;\, \bar{\beta}_{X} \sin \theta_{X} \cos \phi_{X}, 
\;\;\,  \bar{\beta}_{X} \sin \theta_{X} \sin \phi_{X}, 
\;\;\,  \bar{\beta}_{X} \cos \theta_{X}\right)  \,,
\\[3mm]
p_{\ell}^{\mu}
&=&
\frac{1}{2} m_{\ell'} \left(
\alpha_{\ell} , 
- \bar{\beta}_{X} \sin \theta_{X} \cos \phi_{X}, 
- \bar{\beta}_{X} \sin \theta_{X} \sin \phi_{X}, 
- \bar{\beta}_{X} \cos \theta_{X}\right)  \,,
\ena
where the factors 
$\alpha_{X} = 1+ ( m_{X}^{2} - m_{\ell}^{2} )/ m_{\ell'}^2$, 
$\alpha_{\ell^{-}} = 1- ( m_{X}^{2} - m_{\ell}^{2} )/ m_{\ell'}^2$ 
and $\bar{\beta}_{X} 
=\sqrt{1 + ( m_{X}^{2}  - m_{\ell}^{2} )^2/m_{\ell'}^{4}  - 2( m_{X}^{2}  + m_{\ell}^{2} )/m_{\ell'}^{2} } $.
The momentum of the two neutrinos in the rest frame of the momentum $p_{X}$
are parameterized as follows (we assume the neutrinos are massless),
\bea
p_{X}^{\mu}
&=&
m_{X} \left(1, 0, 0, 0  \right) \,,
\\[3mm]
p_{\nu_{\ell'}}^{\mu}
&=&
\frac{1}{2} m_{X}\left(
1, 
\;\;\; \sin \theta_{\nu_{\ell'}} \cos \phi_{\nu_{\ell'}}, 
\;\;\;\, \sin \theta_{\nu_{\ell'}} \sin \phi_{\nu_{\ell'}}, 
\;\;\,\, \cos \theta_{\nu_{\ell'}}  \right) \,,
\\[3mm]
p_{ \bar{\nu}_{\ell} }^{\mu}
&=&
\frac{1}{2} m_{X} \left(
1, 
- \sin \theta_{\nu_{\ell'}} \cos \phi_{\nu_{\ell'}}, 
- \sin \theta_{\nu_{\ell'}} \sin \phi_{\nu_{\ell'}}, 
- \cos \theta_{\nu_{\ell'}}  \right) \,.
\ena

\section*{Acknowledgements}
This study is supported by the Natural Science Basic Research Plan in Shaanxi Province of China under Grant No. 2018JQ1018, and the Innovation Capability Support Program of Shaanxi Province under Grant No. 2021KJXX-47.

\bibliography{Axion}

\end{document}